\documentclass{spie}
 
\usepackage{amsmath,amsfonts,amssymb}
\usepackage{graphicx}
\usepackage[colorlinks=true, allcolors=blue]{hyperref}

\title{Euclid's Near-Infrared Spectrometer and Photometer ready for flight -- review of final performance}

\author[a]{E. Medinaceli}
\author[a]{L. Valenziano}
\author[a]{N. Auricchio}
\author[a]{E. Franceschi}
\author[a]{F. Gianotti}
\author[a]{P. Battaglia}
\author[a]{R. Farinelli}
\author[b]{A. Balestra}
\author[c]{S. Dusini}
\author[c]{C. Sirignano}
\author[c]{E. Borsato}
\author[c]{L. Stanco}
\author[c]{A. Renzi}
\author[c]{A. Troja}
\author[c]{L. Gabarra}
\author[d]{S. Ligori}
\author[d]{V. Capobianco}
\author[d]{L. Corcione}
\author[d]{D. Bonino}
\author[e]{G. Sirri}
\author[e]{L. Patrizii}
\author[e]{M. Tenti}
\author[e]{D. Di Ferdinando}
\author[e]{C. Valieri}
\author[e]{N. Mauri}
\author[f]{F. Giacomini}
\author[g]{D. Le Mignant}
\author[g]{E. Prieto}
\author[g]{M. Carle}
\author[g]{F. Ducret}
\author[h]{W. Gillard}
\author[h]{A. Secroun}
\author[i]{T. Maciaszek}
\author[j]{S. Ferriol}
\author[k]{R. Barbier}
\author[l]{F. Grupp}
\author[m]{W. Holmes}
\author[m]{M. Pniel}
\author[m]{A. Waczynski}
\author[m]{S. Prado}
\author[m]{M. Seiffert}
\author[m]{M. Jhabvala}
\author[n]{RJ. Laureijs}
\author[n]{G. Racca}
\author[n]{JC. Salvignol}
\author[n]{T. Boenke}
\author[n]{P. Strada}
\author[.]{on behalf of the Euclid Consortium}

\affil[a]{INAF-Osservatorio di Astrofisica e Scienza dello Spazio, Via Gobetti 5, Bologna, Italy}
\affil[b]{INAF-OAPd, Vicolo dell'Osservatorio 5, Padova, Italy}
\affil[c]{INFN-Pd, Via marzolo 8, Padova, Italy}
\affil[d]{INAF-OATo, Pino Torinese, Torino, Italy}
\affil[e]{INFN-Bo, Viale B. Pichat 6/2, Bologna, Italy}
\affil[f]{INFN-CNAF, Viale B. Pichat 6/2, Bologna, Italy}
\affil[g]{LAM, 38 Rue Frédéric Joliot Curie, Marseille, France}
\affil[h]{Aix-Marseille Université, CNRS/IN2P3, CPPM, Marseille, France}
\affil[i]{IRAP, 9 Av.\ du Colonel Roche, Toulouse, France}
\affil[j]{IPNL, 4 Rue Enrico Fermi, Villeurbanne Cedex, Lyon, France}
\affil[k]{IP2I, 4 Rue Enrico Fermi, Villeurbanne, France}
\affil[l]{MPE, Gießenbachstraße 1, Garching Baviera, Germany}
\affil[m]{NASA, USA}
\affil[n]{European Space Agency/ESTEC}
\authorinfo{Further author information: (Send correspondence to)\\E-mail: eduardo.medinaceli@inaf.it, Telephone: (+39) 0516398698}

\begin{document} 
\maketitle

\begin{abstract}
ESA's mission \emph{Euclid}, while undertaking its final integration stage, is fully qualified. 
\emph{Euclid} will perform an extra galactic survey ($0<z<2$) by observing in the visible and near-infrared wavelength range. To detect infrared radiation, it is equipped with the Near Infrared Spectrometer and Photometer (NISP) instrument, operating in the 0.9--2\,\textmu m range. In this paper, after introducing the survey strategy, we focus our attention to the NISP Data Processing Unit's Application Software, highlighting the experimental process to obtain the final parametrization of the on-board processing of data produced by the array of 16 Teledyne HAWAII-2RG (HgCdTe) detectors. We report results from the latest ground test campaigns with the flight configuration hardware - complete optical system (Korsh anastigmat telescope), detectors array (0.56 deg$^2$ field of view) and readout systems (16 Digital Control Units and Sidecar ASICs). Performance of the on-board processing is then presented. 
We also describe a major issue found during the final test phase. We show how the problem was identified and solved thanks to an intensive coordinated effort of an independent review `Tiger' team, lead by ESA, and a team of NISP experts from the Euclid Consortium.  An extended PLM level campaign at ambient temperature in Li\`ege and a dedicated test campaign conducted in Marseille on the NISP EQM model eventually confirmed the resolution of the problem.
Finally, we report examples of the outstanding spectrometric (using a Blue and two Red Grisms) and photometric performance of the NISP instrument, as derived from the end-to-end payload module test campaign at FOCAL 5 -- CSL; these results include the photometric Point Spread Function (PSF) determination and the spectroscopic dispersion verification. 
\end{abstract}


\footnote{Presented in Proc. SPIE 12180, Space Telescopes and Instrumentation 2022: Optical, Infrared, and Millimeter Wave, 121801L (27 August 2022)}

\section{INTRODUCTION, THE \emph{EUCLID} MISSION}
\label{sec:intro}  
\emph{Euclid} is an ESA mission that will use an extremely sophisticated space telescope to measure, with unprecedented accuracy, the shape of more than a billion galaxies and the redshifts $z$ of tens of millions of them\cite{redbook}.

\begin{figure} [ht]
   \begin{center}
   \begin{tabular}{c} 
   \includegraphics[height=7 cm]{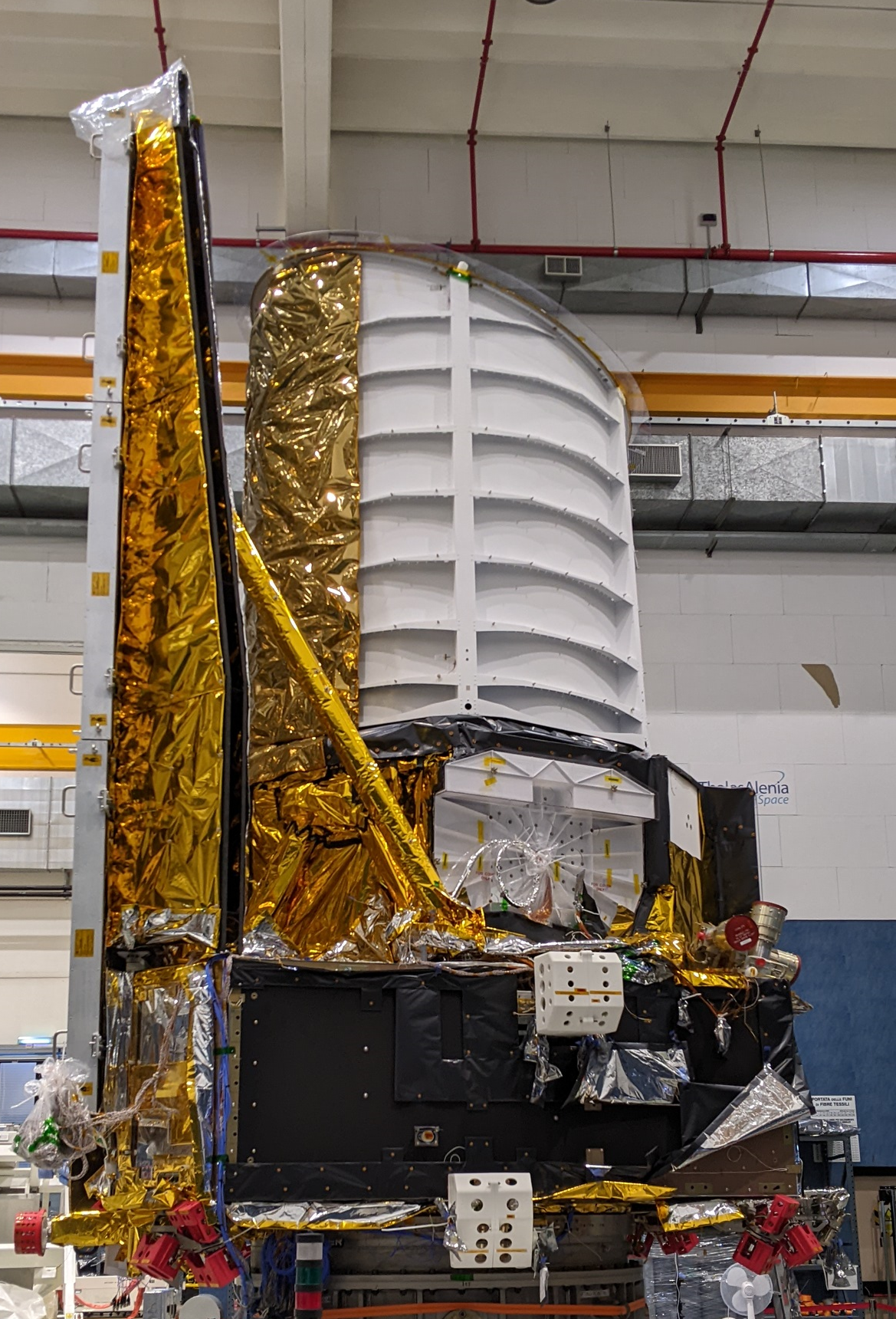}
   \end{tabular}
   \end{center}
   \caption[euclid] 
   { \label{fig:euclid} 
\emph{Euclid}'s fully integrated flight model, picture courtesy of Thales Alenia Space -- Italy.}
   \end{figure}

The mission is developed by the Euclid Consortium and by ESA, with a contribution by NASA. The Spacecraft and the Payload are built by a team of industries, lead by Thales Alenia Space Italia (TAS-I) and Airbus Defence and Space (ADS), respectively. 
The overall mass of the \emph{Euclid} satellite is about 2020\,kg and can be enclosed in a cylinder 3.1\,m in diameter and 4.5\,m high. It is equipped with a telescope with a primary aperture of 1.2\,m, and 2 instruments. The visible imager (VIS) is sensitive to 0.5--0.9\,$\mu$m and the Near-Infrared Spectrometer and Photometer (NISP) is sensitive to 0.9--2\,$\mu$m, both observing a common field of view (FoV). The communication to the ground is provided by two antennas for the two bands: X/X (8.2\,GHz) and K--band (25.5--27\,GHz). The down-link rate to the ground station is in K--band is 55\,Mbits/s, for an average amount of 850\,Gbit transferred daily in the 4\,h  communication interval. Figure \ref{fig:euclid} shows the integrated flight model at TAS-I premises.

\emph{Euclid} will perform a 15\,000\,deg$^{2}$ main survey and a 40\,deg$^{2}$ deep survey in 6.25\,years, observing the sky in the halo orbit around the second Lagrangian (L2) point of Sun-Earth system. 

\section{SCIENCE OBJECTIVES}
\label{sec:science}
The main science objectives are to measure the expansion history $H(z)$ to high accuracy, as to detect percent variations of the dark energy equation of state with robust control of the systematics, i.e. verify the dynamics of the dark matter evolution. \emph{Euclid} will accomplish these objectives using two probes: the scale of Baryonic Acoustic Oscillations (BAO) in the clustering pattern of galaxies as a standard rod; by measuring the spectroscopic redshifts of 50 million galaxies in the redshift range
0.7\,$<z<$\,2.1 the three-dimensional galaxy distribution of the Universe can be mapped to high precision\cite{redbook}. And using shape distortions induced by weak gravitational lensing; by measuring the correlations in the shapes of the 1.5 billion galaxies, the expansion and growth history of the Universe can be determined with high accuracy\cite{redbook}. 

Another objective is to measure the growth rate of structures from two independent probes: clustering redshift-space distortions and weak lensing tomography. These probes will allow us to detect possible deviations from General Relativity with very high sensitivity\cite{redbook}.
\section{\emph{EUCLID} SURVEY}
\label{sec:survey} 
\begin{figure} [htb]
   \begin{center}
   \begin{tabular}{c}
   \includegraphics[height=5.5 cm]{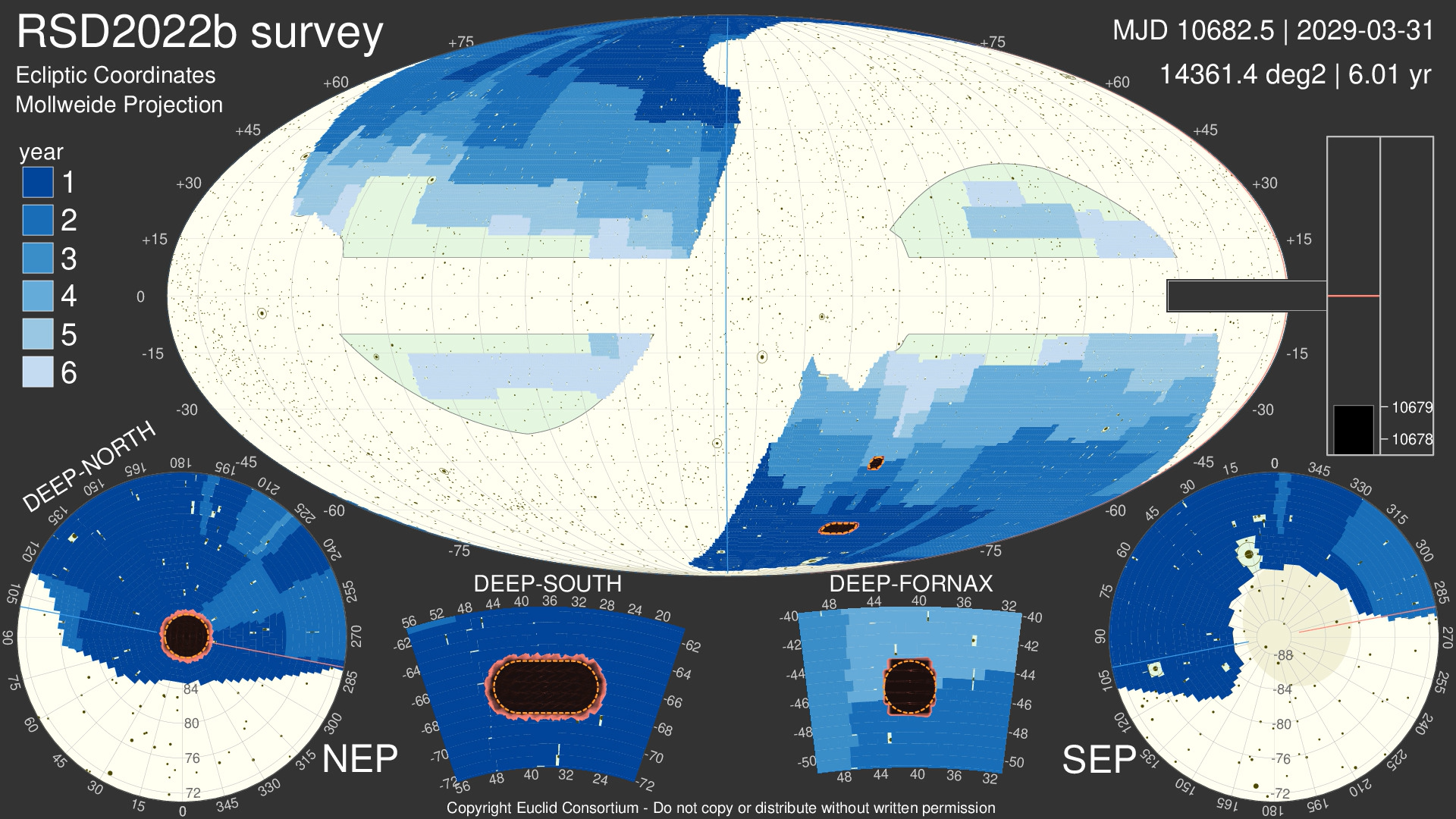}
   \end{tabular}
   \end{center}
   \caption[survey] 
   { \label{fig:survey} 
  \emph{Euclid}'s survey (version RSD2020b of 12/31/2022) after 6 years of mission lifetime represented in ecliptic coordinates using the Mollweide projection, where the Wide Survey (15\,000\,deg$^2$) is plotted using a blue color scale. We also show the 4 regions (NEP, DEEP-SOUTH, FORNAX and SEP) of the Deep Survey ($\sim$50\,deg$^2$). The exclusion zones are reported in white. Image courtesy of J. Dinis - \emph{Faculdade de Ciências da Universidade de Lisboa}.
  }  
   \end{figure}
\emph{Euclid}'s survey strategy is based in a step-and-stare mode,  with a FoV around $0.5$\,{\rm deg$^2$}. The line spacecraft-Sun moves 1\,degree per day over the ecliptic plane, and every day the sky area covered is 
15--20 degrees, depending on the daily scanning strategy.
Fig.\ref{fig:survey} shows \emph{Euclid}'s global Wide Survey (represented by a blue color scale) coverage of 15\,000\,{\rm deg$^2$} of the extragalactic sky, which is more than 35$\%$ of the celestial sphere, in more than 40\,000 observations obtained during the mission lifetime of approximately 6 years. 
\emph{Euclid} will also perform a Deep Survey, approximately 2 magnitudes fainter than the Wide-survey, over an area about $50$\,{\rm deg$^2$}, split in four regions of 10--20\,deg$^2$ (DEEP-NORTH, DEEP-SOUTH, DEEP-FORNAX, and SEP) shown in the same figure. The exclusion zones are $\pm25^{\circ}$ from the Galactic plane latitude, to avoid stellar contamination, and $\pm15^{\circ}$ from the ecliptic plane to avoid zodiacal light contamination, represented in the same figure as white regions.
\emph{Euclid} will measure the shapes of 50\,000 galaxies per field, for a total of 1.5\,billion galaxies and the spectra of 850\,galaxies/field leading to a total of 25\,million galaxy spectra. More details can be found in a dedicated paper.  \cite{Euclid:2021icp}

\subsection{The EUCLID WIDE SURVEY}
\label{sec:widesurvey}
A \emph{panchromatic view} strategy is used for the observations of the same FoV using complementary measurements in the visible band (with the VIS detector) and with both photometric and spectrometric images in the infrared band (with the NISP detector). We report here an illustration on how the sequence of observations is built: the NISP application software is \emph{flexibly} designed to implement this strategy. The dithering technique is used to correct instrumental effects on images. The \emph{Euclid} Reference Observing Sequence (ROS) is the elementary building block of the Euclid Wide Survey. The ROS is composed of 4 smaller sequences of operations, called \emph{Dither} in the \emph{Euclid} jargon.
Each \emph{Dither} is made of a VIS exposure and of a single NISP spectrometric exposure (done in parallel), followed by 3 photometric exposures. A small pointing change (dither slew) concludes each \emph{Dither}. The $4^{th}$ \emph{Dither} includes a NISP dark photometric exposure at the end. After each ROS, the telescope is moved to a new pointing (field slew).
The ROS duration, which includes also the time of dither slews but not the field slew, is 4214\,s. Approximately 20 ROS are completed every day.  

A complete NISP ROS is build as follows, where the Filter Wheel position is indicated with FWA, the Grism Wheel position with GWA and the exposure number with Exp\#:

\begin{center}
\begin{tabular}{ |c|c|c|c|c|c| } 
  \hline
Dither & Exp 1 & Exp 2 & Exp 3 & Exp 4 & Exp 5 \\
  \hline
 & (FWA, GWA) & (FWA, GWA) & (FWA, GWA) & (FWA, GWA) & (FWA, GWA)\\
  \hline
1 & Open, RGS0 & J$_{\mathrm{E}}$, Open & H$_{\mathrm{E}}$, Open & Y$_{\mathrm{E}}$, Open & ---\\ 
  \hline
2 & Open, RG180+4 & J$_{\mathrm{E}}$, Open & H$_{\mathrm{E}}$, Open & Y$_{\mathrm{E}}$, Open & ---\\ 
  \hline
3 & Open, RG0-4 & J$_{\mathrm{E}}$, Open & H$_{\mathrm{E}}$, Open & Y$_{\mathrm{E}}$, Open & ---\\
  \hline
4 & Open, RG180 & J$_{\mathrm{E}}$, Open & H$_{\mathrm{E}}$, Open & Y$_{\mathrm{E}}$, Open &  Closed, RGS180\\ 
 \hline
\end{tabular}
\end{center}


\section{\emph{EUCLID} INSTRUMENTATION}
\label{sec:instrumentation}
\emph{Euclid}'s complete telescope (structure and mirrors) is made of sintered silicon carbide (SiC), which combines a high stiffness, low density, and high thermal conductivity, thus reducing thermal gradients along and across the optical path. The telescope has a three--mirror Korsch configuration with 0.45\,{\rm deg} off-axis field with 1.2\,{\rm m} aperture of the primary mirror, and a secondary mirror of 0.35\,{\rm m}. Together providing a collecting area of 1\,{\rm m$^{2}$}. Visible and near-infrared observations are done with the same FoV using a dichroic beam-splitter \cite{telescope}.

The VIS instrument (spectral range of 550--900\,{\rm nm}) focal plane is composed of 6$\times$6 CCDs (Teledyne e2v, 12$\times$12\,{\rm $\mu$m$^{2}$}\,pixels, 4096$\times$4096 pixels$^2$) with a resolution of 0.1\,arcsec/px and a FoV = 0.787$\times$0.709\,deg$^2$ with a focal length of 24.5\,m. Its limiting magnitude is 24.5 for extended sources at 10$\sigma$. The VIS data rate is less than 520\,Gb/day.

The NISP (spectral range of 900--2000\,{\rm nm}) focal plane is composed of 4$\times$4 H2RG (HgCdTe) CMOS (Teledyne H2RG, 18$\times$18$\mu$m$^{2}$ pixels, 2048$\times$2048\,pixels$^2$) with a resolution 0.3\,arcsec/px and a FoV = 0.55\,deg$^2$ with a focal length of $\sim$6.1\,m. NISP's data rate is $\sim$290\,Gb/day. For the photometry, the limiting magnitude is 24, the accuracy of mean redshift of each tomographic bin is better than $0.002\,(1 + z)$ (see \emph{Euclid} requirements\cite{redbook}) and the spectral range is $937\,-\,2057$\,nm (0.1\% cut-on and cut-off) \cite{Schirmer}. 
For the spectroscopy, the limiting magnitude is 19.5 and the redshift accuracy is better than $0.001\,(1 + z)$, as per \emph{Euclid} requirements\cite{redbook}. The spectral range is $920\,-\,1950$\,nm\cite{Schirmer}, where the red-shifted $H_{\alpha}$ (656 nm) can be observed in the interval $z$=0.8--1.8. 

This paper presents the main features and performance of the NISP detector, including the on-board signal pre-processing.

\subsection{THE NISP INSTRUMENT}
\label{sec:nisp}
\begin{figure} [ht]
   \begin{center}
   \begin{tabular}{c} 
   \includegraphics[height=5.5 cm]{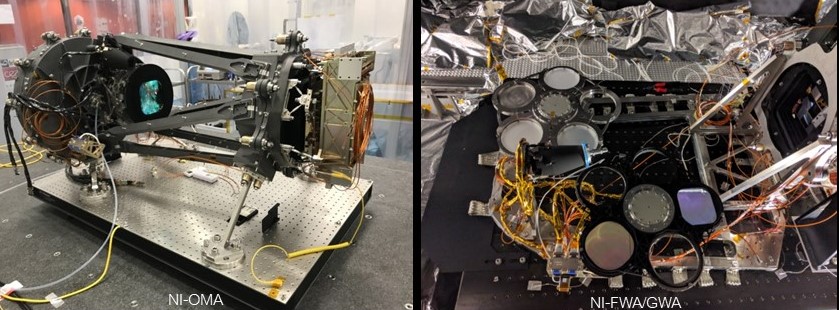}
   \end{tabular}
   \end{center}
   \caption[nisp] 
   { \label{fig:nisp} 
\emph{Left:} NISP detector in its final configuration. Only the CaLA  is visible (the CoLA is on the other side of the wheels' enclosure); the wheels' enclosure can be seen on the far left, and the focal plane on the far right. The complete system is sustained by a SiC structure. \emph{Right:} the filter wheel (top) and the grism wheel (bottom) in a horizontal position during a ground test at LAM. The calibration unit is also visible at the center of the image.}
   \end{figure}

\begin{figure} [ht]
   \begin{center}
   \begin{tabular}{c} 
   \includegraphics[height=5.5 cm]{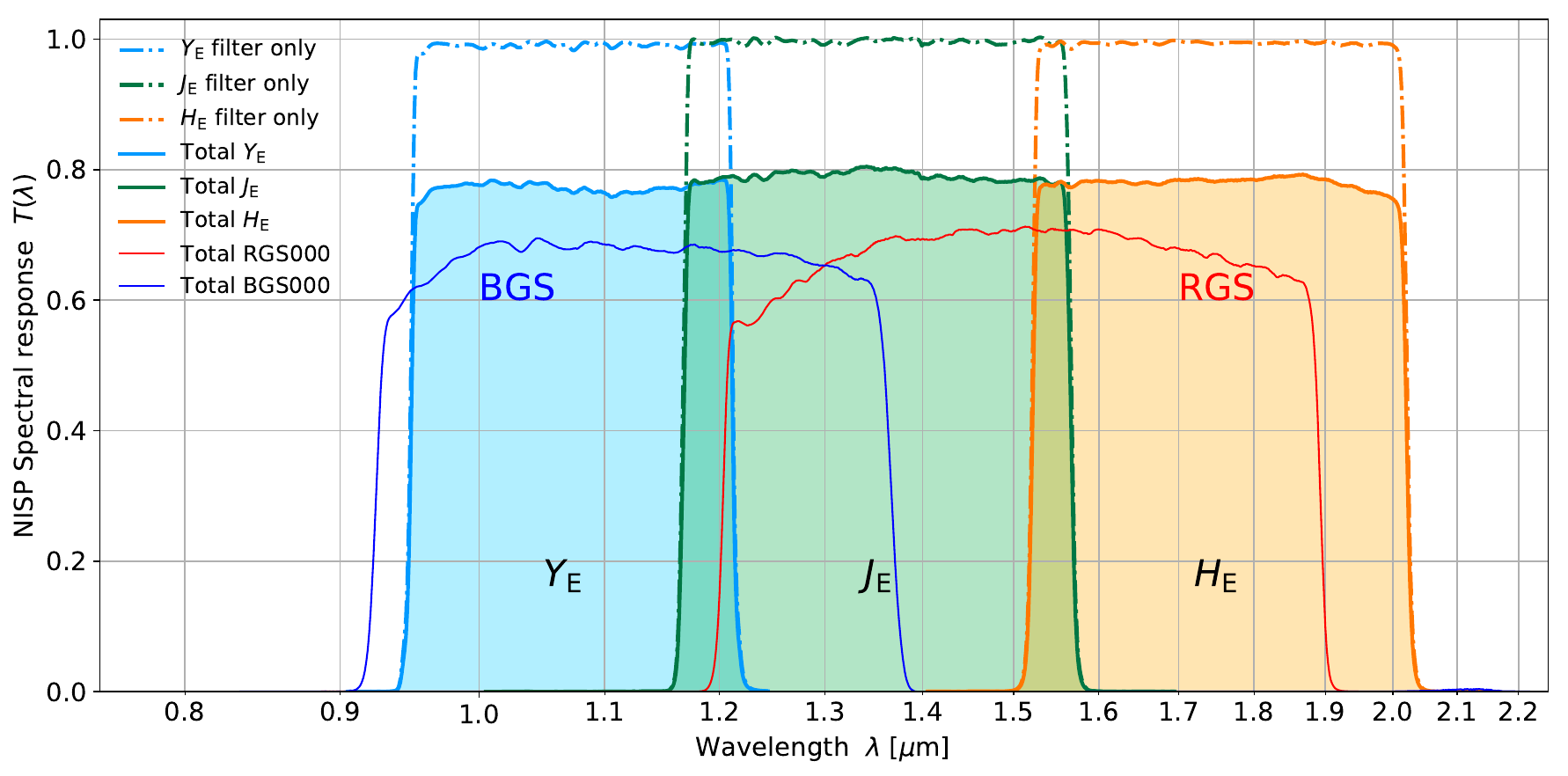}
   \end{tabular}
   \end{center}
   \caption[passbands] 
   { \label{fig:passbands} 
   NISP transmission elements. The thin lines show the effective filter transmission integrated over the beam footprint, and the shaded areas the total response. Total RGS000 and BGS000 grisms spectral response is plotted with solid lines \cite{Schirmer}.
   }
   \end{figure}

The \emph{Euclid} NISP instrument comprises an opto-mechanical assembly (NI-OMA) and an Infrared Detector Assembly (NI-DA). The NI-OMA includes a collimator (CoLA) and a focusing (CaLA) lens assembly, mounted on a SiC structure, and the NI-DA that accommodates the infrared detectors.
Between CoLA and CaLA, two wheels equipped with filters and grisms provide NISP with spectrometric and photometric capabilities. Fig.\ref{fig:nisp} shows the components of the NI-OMA. On the left-side of the figure, the optical and mechanical support assemblies are shown; on the right-side, the filter (FWA) and grism (GWA) wheel systems are shown. For the photometry NISP uses three filter passbands: Y$_E$\,(949.6--1212.3\,nm\,$\pm$\,0.8\,nm), J$_E$\,(1167.6--1567.0\,nm\,$\pm$\,0.8\,nm), and H$_E$\,(1521.5--2021.4\,nm\,$\pm$\,0.8\,nm) (50\% cut-on and cut-off, and $\sigma_{50}$ uncertainly)\cite{Schirmer} represented in Fig.\ref{fig:passbands}. While for the slitless spectroscopy NISP uses 4 grisms: one blue (920--1400 nm) with single orientation of $0^{\circ}$, and 3 red (1200--1950\,nm) \cite{Schirmer} with orientations $0^{\circ}$, $90^{\circ}$ and $180^{\circ}$, also represented in the same Fig.\ref{fig:passbands}. 

The NISP detector system is composed of an array of 16 HAWAII-2RG sensors from Teledyne Imaging Scientific, accomodated in a 4$\times$4 configuration, with $\sim$90\,K operating temperature. A molybdenum mechanical structure holds the detectors and a burnished aluminum truss holds the 16 SIDECAR ASICs (operating at $\sim$135\,K). This structure can be seen at the right-side in both images of Fig.\ref{fig:nisp}. In the same figure, right panel, the focal plane array is partially visible on the right-side. 

\begin{figure} [ht]
   \begin{center}
   \begin{tabular}{c} 
   \includegraphics[height=5.5 cm]{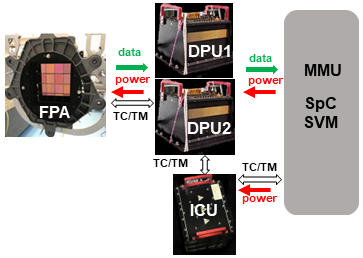}
   \end{tabular}
   \end{center}
   \caption[we] 
   { \label{fig:we} 
   Scheme of NISP main communication and power interfaces. The NISP warm electronics is composed of two identical Data Processing Units (DPU1 and DPU2 shown at the center of the scheme). They control the NISP focal plane (FPA, shown at the left-side). DPUs are commanded by the Instrument Control Unit (ICU shown at the bottom) that is interfaced with the Spacecraft's Service Module (SpC SVM, represented on the right-side of the scheme). White arrows represents the commandability and telemetry retrieval flux - using a MILBUS1553 link, green arrows shows the science data flux stored on the SpC's Mass Memory Unit (MMU), using a SpaceWire interface. Red arrows represent the power distribution provided by the SpC. A picture of the NI-WE components can be seen also on the left-side of Fig.\ref{fig:tests}.}
   \end{figure}

The detector system is controlled by two identical Data Processing Units (DPU), performing synchronous operations, each one controlling half of the focal plane. Each DPU is equipped with 8 Digital Control Units (DCUs), each one interfaced to an ASIC. The two DPU are commanded by the Instrument Control Unit (ICU) that receives telecommands from the spacecraft. While science data are directly sent to the Mass Memory Unit, the ICU collects the DPU telemetry, the ICU, packetized and sent to the spacecraft on-board computer. The ICU monitors some selected DPU parameters to trigger instrument level functions, called FDIR (Failure Detection Isolation and Recovery) and other events (e.g. end of exposure, end of transmission). The ICU and the two DPU together compose the so-called \emph{warm electronics} and are hosted by the satellite Service Module (SVM) that operates at around 293\,K. This is schematically illustrated in Fig.\ref{fig:we}. 

\subsection{NISP ON-BOARD PROCESSING}
\label{sec:asw}
\begin{figure} [ht]
   \begin{center}
   \begin{tabular}{c} 
   \includegraphics[height=5.5 cm]{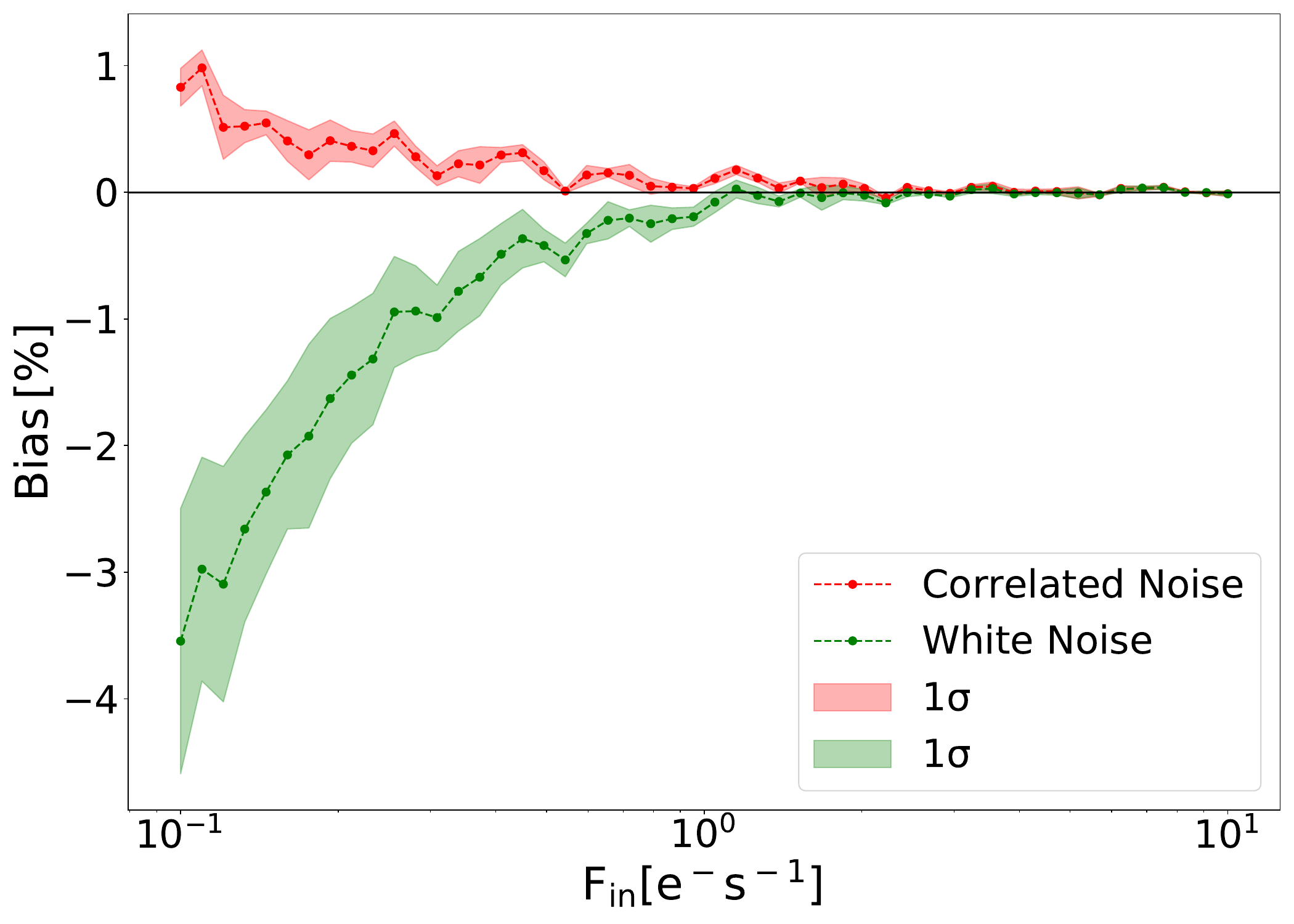}
   \end{tabular}
   \end{center}
   \caption[noise] 
   { \label{fig:noise} 
   Reconstructed flux estimation bias as a function of the flux. Two approximations are shown, correlated noise (pink) and white noise (green). Both models provides an almost null ($\sim$0.1$\%$) bias for a noise flux in the interval between 1 and 2 $e^-$ s$^{-1}$, typically expected for \emph{Euclid} observations \cite{jimenez}.}
   \end{figure}
The DPU application software (ASW) architecture is driven by the science performance, operating and controlling dedicated electronics suited to cope with the time constraints of the NISP acquisition sequences during the sky survey. The same DPU-ASW running in both DPUs implements a multi-task preemptive scheduling algorithm written in ansi-C (ECSS standards) running in a VxWorks 5.1 real-time operative system. Using 13 tasks, it handles the spacecraft's time distribution and the synchronization of all the subsystems at a frequency of 1\,Hz, a command rate of approximately 512\,bit (1\,Hz), a telemetry rate of around 18\,kbits (40\,Hz), and a science data production around 290\,Gbit per day.  
The main task of the DPU-ASW is to implement the on-board pipe-line, allowing all basic operations required to evaluate the signal acquired using the multi-accumulation charge collection (\emph{MACC}) technique. The processing includes averaging groups of the up-the-ramp input exposure frames, where the signal estimation is obtained with a weighted least-square fit of the accumulated charge derivatives. A quality statistical estimator of the fit is evaluated ($\chi^{2}$, with different resolutions for each exposure mode i.e. spectrometric using 16\,bit/px or photometric using 1\,bit/px). The on-board processing algorithm is executed recursively for all 16 detectors (8 per DPU), considering for each detector every single sensitive pixel (2040$\times$2040\,px$^{2}$). A subset of 
32\,704\,px (external frame of 4 pixels of the 2048$\times$2048\,px$^{2}$ array) 
is used to correct the image for the thermal background noise. The DPU-ASW performs a data-volume compression before the ground down-link \cite{proc}.
\begin{figure} [htb]
   \begin{center}
   \begin{tabular}{c} 
   \includegraphics[height=5 cm]{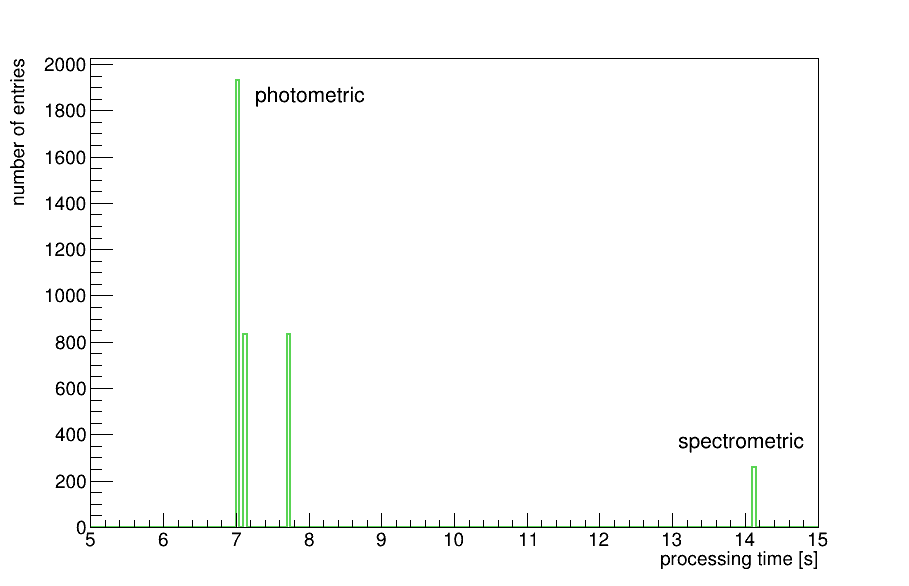}
   \end{tabular}
   \end{center}
   \caption[proctime] 
   { \label{fig:proctime}
   NISP Data Processing Unit performance - on-board processing time. The time depends on the acquisition type \emph{i.e.}\ spectrometric (spectro) using 15 co-added frames for the linear fit, or photometric (photo) using 4 groups. The mean processing duration intervals (estimated on the Thermo-Vacuum TV3 dataset) are approximately 7.1\,s and 14.1\,s for photometric and spectrometric acquisitions, respectively. Both intervals have standard deviations below 0.1\,s. The second peak in the photometric distribution comes from data with a processing test-configuration.
   }
   \end{figure}
The on-board processing algorithm depends on two main parameters (personalized for each detector): the detector's read-out-noise (RON) and the gain. The correlated RON, which includes the Poissonian photon noise, can be approximated as a white noise \cite{kubik} and its distribution can be assumed to be gaussian. This paper\cite{jimenez} reports that for \emph{Euclid}'s sky observations (RON in the range between 1 and 2\,$e^-$ s$^{-1}$) the gaussian assumption, as implemented in the on-board processing, introduces a bias lower than approximately 0.1\,$\%$, as shown in Fig.\ref{fig:noise}. The detector gain may be considered as the combination of two terms: the charge-to-voltage conversion (V/$e^-$) and the analog-to-digital conversion (ADU/V).
   \begin{figure} [htb]
   \begin{center}
   \begin{tabular}{c} 
   \includegraphics[height=6 cm]{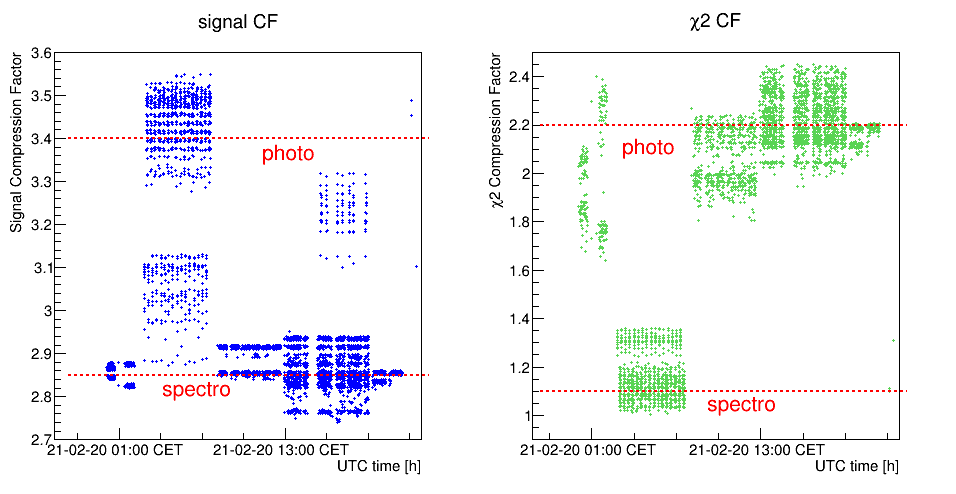}
   \end{tabular}
   \end{center}
   \caption[compfactors] 
   { \label{fig:processing} 
   Data Processing Unit on-board performance. Compression Factor (CF) of $\sim$4000 processed images and $\chi^{2}$ frames. Results are different for spectrometric (labelled spectro) and photometric (labelled photo) datasets, the horizontal lines indicates the mean values. \emph{Left-side}: science frames results: spectro CF = 2.85, photo CF = 3.4. \emph{Right-side}: $\chi^{2}$ frames results: spectro $\chi^{2}$ CF = 1.15 , photo $\chi^{2}$ CF = 2.2.}
   \end{figure}
The RON and the gain for each detector were derived experimentally during dedicated characterization test campaigns (TV1/TV2, see Sect.\ref{sec:tests}). The following data processing performance results are obtained using the final detector's parametrization. The processing time of spectrometric and photometric acquisitions were derived from the data of the latest test campaigns, being approximately 7.1\,s and 14.1\,s for photometric and spectrometric acquisitions, respectively, as in Fig.\ref{fig:proctime}.  The size of the NISP data products depends on the results of the on-board compression algorithm; in \emph{Euclid} NISP, a loss-less compression (modified NASA CFITSIO, based on the Rice algorithm\cite{proc}) is implemented in the DPU-ASW. Its Compression Factor (CF, defined as the ratio of the original to the compressed frame sizes) depends on the acquisition type. Fig.\ref{fig:processing} reports the signal (\emph{left}) and $\chi^{2}$ (\emph{right}) CF obtained during NISP performance tests. In the plot, the mean values CF$_\mathrm {spectro~science}$ = 2.85, CF$_\mathrm{photo~science}$ = 3.4, CF$_\mathrm{spectro~\chi^2}$ = 1.15 and CF$_\mathrm{spectro~\chi^2}$ = 2.2 are indicated with dashed horizontal lines. 
Each detector compressed data product, transmitted individually to the MMU, is composed of the science frame, the $\chi^{2}$ frame, a header, and a set of telemetry. The average data size of a nominal NISP single \emph{Dither} (full FPA), see Sect.\ref{sec:widesurvey}, is 379.2\,MByte. Assuming 20 ROS per day, approximately 26.3\,GByte is the average data volume produced per day.
\section{NISP TEST CAMPAIGNS}
\label{sec:tests}  
\begin{figure} [ht]
   \begin{center}
   \begin{tabular}{c} 
   \includegraphics[height=5.5 cm]{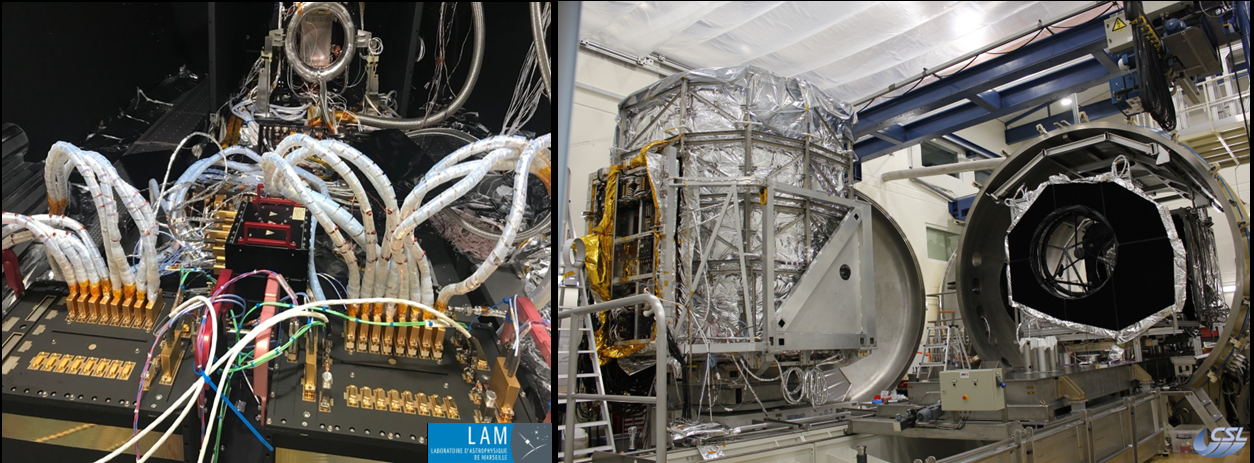}
   \end{tabular}
   \end{center}
   \caption[campaigns] 
   { \label{fig:tests} 
   \emph{Left}: NISP tests setup at \emph{Laboratoire d’Astrophysique di Marseille}, France. NISP warm electronics close-up, from bottom to top: the two DPUs, the ICU, and the focal plane array. \emph{Right}: \emph{Euclid}'s end-to-end PLM test setup at the \emph{Centre Spatial de Liège}, Belgium. The NISP warm electronics mounted on the panel Y can be seen at the left-side of the picture (vertical structure partially covered with MLI), and the telescope placed horizontally at the right-side of the image.}
   \end{figure}

The NISP instrument was extensively tested at the \emph{Laboratoire d’Astrophysique di Marseille (LAM)} France, in different sessions during the Flight Model ground test campaign; the test setup is shown in the left-side of Fig.\ref{fig:tests}. During two  sessions, called TV1 and TV2 of the thermal vacuum and thermal balance (TBTV) test campaign, we fully characterized the detectors  and we obtained an individual set of parameters to optimize the throughput of each single detector. Afterwards, in the TV3 test session, we measured the performance of the optical and detector systems. The following approximated figures describe the outcome of this session, which lasted approximately 3 months: 40\,K commands were executed, 19\,K images per detector were acquired, 300\,K files were produced, and a continuous run of 72\,h of wide survey ROS was exercised.
In this test campaign, we measured the accurate position of the NISP focus, as well as the instrument's dark level. We also evaluated the grisms' (spectrometric) and filters' (photometric) optical quality, and the spectral resolution by testing the dispersion for the red grisms RGS000, RGS180 and blue grism BGS000. We verified the requirements of the signal-to-noise ratio for the spectroscopy with RGS000, RGS180, and BGS000, and for the photometry with Y$_E$, J$_E$ and H$_E$ bands\footnote{more details in Paper 12180-57 of this proceedings}.

In 2021, during the \emph{Euclid} end-to-end Payload Module (PLM) test campaign, both NISP and VIS instruments were integrated with the telescope, and their global performance was derived. The campaign was carried out at the \emph{Centre Spatial de Liège} (CSL), Belgium by Airbus Defense and Space (ADS) in one month of nominal continuous operations in flight-representative operating conditions. The setup can be seen in the picture on the right of Fig.\ref{fig:tests}.
The NISP/VIS auto-compatibility (cross talk) was tested, demonstrating that the instruments are not mutually interfering, and the NISP/VIS common focus position with the telescope was obtained. For NISP the reference dark level was obtained, as well as the photometric and spectrometric point spread functions (PSF), and the spectroscopic dispersion was verified. NISP's photometric PSF and the spectrometric dispersion are presented in Sect.\ref{sec:performances}.
\subsection{Major issue during PLM test campaign}
\label{sec:fifo} 
Unexpectedtly, only 7 out of 16 NISP detectors worked correctly during the CSL campaign at cryogenic conditions (FPA temperature $\sim$90K). We observed \emph{artifacts} in the science images of the 9 non-working detectors, such as horizontal or vertical patterns Moreover, all 9 detectors stopped producing data after the first exposure of each sequence. This anomalous behavior was not present at CSL during the reference tests at room temperature using the same hardware configuration, nor on the cryogenic tests at LAM, using the same flight hardware.
Despite this limitation, \emph{Euclid}'s optical performance (listed in Sect.\ref{sec:tests}) could be verified using correctly operating detectors only, luckily located both at the edge and at the center of the NISP FPA.

\begin{figure} [ht]
   \begin{center}
   \begin{tabular}{c}
   \includegraphics[height=3.2 cm]{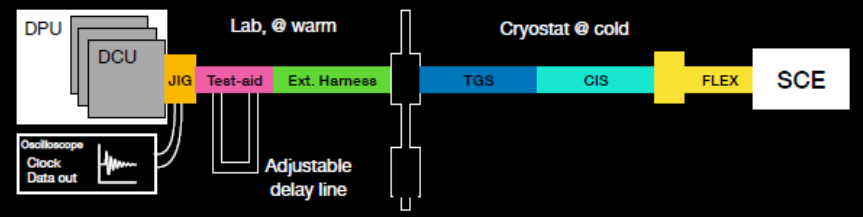}
   \end{tabular}
   \end{center}
   \caption[tv4] 
   { \label{fig:tv4} 
   The cryogenic test setup used at LAM. The different segments (inside and outside the cryostat) of the connection harness between the DCU and the ASIC (SCE) are shown. The test aid (delay line) allows to inject an adjustable delay in the signal propagation, and digital signals were accessible through the break-out box (JIG).
}
   \end{figure}

A large troubleshooting strategy was put in place to address this issue. This included an external ESA team of experts to scrutinize the NISP operations, brainstorming with the detector and readout electronics provider (NASA) and DPU hardware provider (OHB-I). Industrial partners (TASI and ADS) performed independent analyses and looked for possible external factors coming from the test facility. The only difference found at the CSL setup with respect to LAM setup was the introduction of extension harnesses connecting the FPA with the DCUs of the DPUs. To characterize the electrical signals between the front-end electronics (sidecar ASICs) and the Digital Control Units (details in Sect.\ref{sec:nisp}), we executed two new dedicated test campaigns to reproduce the detector failure conditions. 
The first campaign was an extension of the PLM tests at warm ambient conditions at CSL. The data there was used as a reference because in these conditions there were no errors generated. 
Then, we built a completely new cryogenic test campaign at LAM, using the Electrical Qualification Models of the WE.  Signal electrical characterization at LAM was done using the hardware configuration shown in Fig.\ref{fig:tv4}: the nominal components of the connection harness between a single DCU and an ASIC (labeled SCE - Sensor Chip Electronics) is composed of different segments in the cold section inside the thermal vacuum cryostat (TGS, CIS, and Flex), as shown in the figure; the harness elements outside the cryostat (at room temperature) are the extension harness and an adjustable delay line (labeled test-aid). Electrical measurements were done by accessing the signal through an \emph{ad-hoc} break-out box (JIG).   
Differences in the transmission time for some detectors in the CSL configuration of the order of some nanoseconds were found. Then, the CSL error condition was reproduced varying the cable's length by some centimeters that introduced a delay in the signal propagation, generating errors and the stopping the ROS nominal operations. The width of the delay window causing the error condition was 3\,ns. The ESA team pointed out a non-correct treatment of these errors. 
Analyzing the induced errors, an inconsistency in the hardware documentation leading to an incorrect DCU-ASIC I/F error handling, as implemented in the DPU-ASW, was found. More specifically, the native double sampling strategy of ASIC signals (using two FIFOs) implemented in the DCU drivers was injecting false positive errors when the signal was sampled outside an acceptance window of one of the two FIFOs, even if the data transmission was correctly recovered by the other FIFO. The \emph{false positives} were interpreted by the software as true errors, and all the science interface was then reset, stopping of the data production. The issue was corrected by implementing a new comprehensive error strategy handling the DCU-ASIC interface in the DPU application software. Further tests confirmed the robustness of the implemented solution.
\section{EXAMPLES OF NISP PERFORMANCE}
\label{sec:performances}  
Some of the NISP performance results obtained during the system-level tests done at the facilities described in Sect.\ref{sec:tests} are presented. 
Other results can be found in article number 12180-57 of this proceedings.
\subsection{NISP Photometric PSF evolution}
\label{sec:psf} 
\begin{figure} [ht]
   \begin{center}
   \begin{tabular}{c}
   \includegraphics[height=3.8 cm]{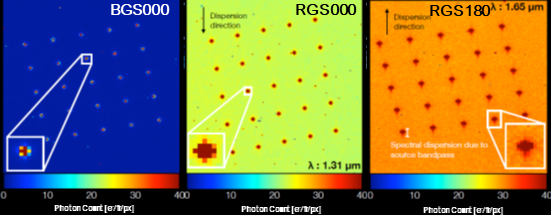}
   \end{tabular}
   \end{center}
   \caption[psf] 
   { \label{fig:psf} 
   NISP photometric PSF determination. Images show the 25\,PSF acquisitions done with the M2M mirror at NISP/VIS best focus position during the CSL test campaign. The image on the left side corresponds to the blue grism BGS000, while the images on the center and on the right side are for the red grism with the RGS000 and RGS180 orientations respectively. 
}  
   \end{figure}

The PSF determination was done using a single long exposure, and because of NISP's sub-pixel size of the PSF, 25\,PSF flashes (grid of 5$\times$5) at different locations were done using a monochromatic point-like source. Figure\ref{fig:psf} shows the photon counts obtained for the blue grism BGS000 (on the left side), for the red grism RGS000 (center), and for the RGS180 (right side). The collimator was at warm temperature, therefore it was emitting thermal radiation, and the background presents special features at a small scale as its flux increases with the wavelength. This introduced a large background during the PSF characterization for the red grism at CSL. This effect was included in the model of the PSF determination. 
\begin{figure} [htb]
   \begin{center}
   \begin{tabular}{c}
   \includegraphics[height=7.3 cm]{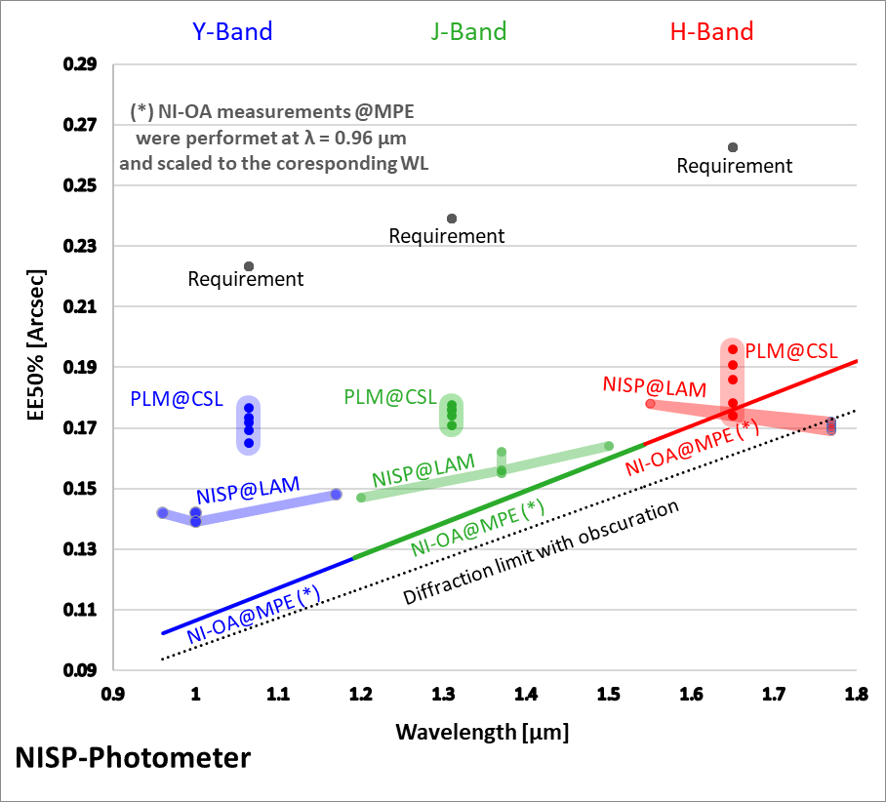}
   \end{tabular}
   \end{center}
   \caption[psfevolution] 
   { \label{fig:psfevolution} 
   Evolution of the diffractive EE50 radius vs. wavelength. NISP (LAM) and end-to-end (CSL) results for the Y$_E$, J$_E$, and H$_E$ bands are plotted with blue, green, and red markers respectively. Stable values are well below the requirements (with black markers) and are near the diffraction limit (dashed line). Reference measurements done with NISP optical system stand-alone (NISP-OA@MPE) are also included.
}  
   \end{figure}
The PSF is modeled by a 2D asymmetric Erf function considering a Gaussian pixel integration, and the 50$\%$ Encircled Energy radius EE50 was deduced from the width of the Gaussian (centroid position of the PSF that contains the 50$\%$ of the total PSF energy). CSL results are compared with the ones obtained during NISP tests at LAM in Fig.\ref{fig:psfevolution}. The figure shows that the EE50 for the different photometric bands is very stable, with median/mean values below the requirement ($<40\%$) indicated with black dots, and close to the theoretical limit (indicated with the dashed line - diffraction limit with obscuration).
\subsection{NISP Spectrometric dispersion verification}
\label{sec:dispersion} 
\begin{figure} [ht]
   \begin{center}
   \begin{tabular}{c}
   \includegraphics[height=0.82 cm]{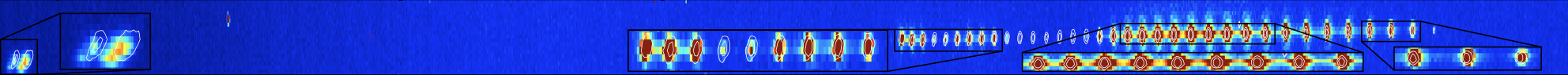}
   \end{tabular}
   \end{center}
   \caption[dispersion] 
   { \label{fig:dispersion} 
   Red-grism RGS000's Fabry--Perot peak position comparison of the spectra obtained at the NISP TV3 tests (LAM), and the end-to-end tests (CSL). TV3 spectra are plotted with a contour scale, while PLM spectra are with a color scale. An almost perfect match was obtained, on the \emph{left-side} the 0$^\mathrm{th}$ order is shown, and on the \emph{rigth-side} the 1$^\mathrm{st}$ order. Inside the black boxes the zoom-in of parts of the spectra are shown.
}  
   \end{figure}

\begin{figure} [ht]
   \begin{center}
   \begin{tabular}{c}
   \includegraphics[height=6.0 cm]{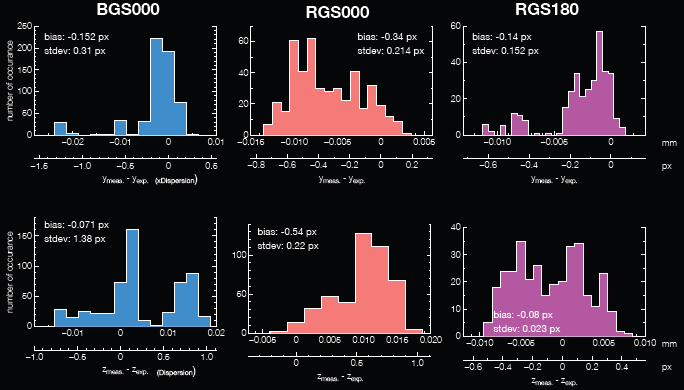}
   \end{tabular}
   \end{center}
   \caption[histo] 
   { \label{fig:histo} 
  Comparison of the distance-dispersion of the TV3 spectra (top row: y$_\mathrm{exp}$ and bottom row: z$_\mathrm{exp}$ bottom row) with the PLM ones. The mean values (bias) and the standard deviation (stdev) are presented for the BGS000, RGS000, and RGS180 grisms. The deviations are within the requirement ($<0.8$)\,px.
  }  
   \end{figure}

The spectrometric calibration was verified by comparing the results from the NISP TV3 test campaign at LAM to the end-to-end ones obtained at CSL. The NISP Optical Ground Equipment used during TV3 was used to verify the spectral dispersion at the PLM level.
During both test campaigns, the same spectral source was used, i.e. etalon. The etalon is a 55\,$\mu$m air-space Fabry--Perot interferometer providing 34 transmission peaks within the NISP's red-grism band pass.
Figure \ref{fig:dispersion} shows the comparison between the two spectra obtained with the RGS000. For the sake of comparison, no rotation or stretching was applied: only a rigid translation of the 0$^\mathrm{th}$ order was added, to visually align them. A perfect match between the RGS000 spectra from TV3 (contour) and PLM spectra (color) is readily evident in the figure.  
The overall dispersion-distance between the Fabry--Perot peak position measured during TV3 and PLM test are well below ($40\%$) the requirement, amounting to $<$0.8\,px. This result shows no evidence of any significant impact of dichroic on the dispersion solution. The results for the projection of the focal plane axis $y$ and $z$ for the BGS000, RGS000 and RGS180 are shown in Fig.\ref{fig:histo}; for each case, the mean values (bias) and the standard deviation are quoted. 
\newpage
\bibliography{report} 
\bibliographystyle{spiebib} 
\end{document}